\documentclass[twocolumn,english]{revtex4-1}
\usepackage[T1]{fontenc}
\usepackage[latin9]{inputenc}
\usepackage{mathtools}
\usepackage{amsmath}
\usepackage{amssymb}
\usepackage{graphicx}
\usepackage{babel}
\begin{document}

\title{Enzyme kinetics at the molecular level}

\author{Arti Dua}

\address{Department of Chemistry, Indian Institute of Technology, Madras,
Chennai-600036, India}
\begin{abstract}
The celebrated Michaelis-Menten (MM) expression provides a fundamental
relation between the rate of enzyme catalysis and substrate concentration.
The validity of this classical expression is, however, restricted
to macroscopic amounts of enzymes and substrates and, thus, to processes
with negligible fluctuations. Recent experiments have measured fluctuations
in the catalytic rate to reveal that the MM equation, though valid
for bulk amounts, is not obeyed at the molecular level. In this mini-review,
we show how new statistical measures of fluctuations in the catalytic
rate identify a regime in which the MM equation is always violated.
This regime, characterized by temporal correlations between enzymatic
turnovers, is absent for a single enzyme and unobservably short in
the classical limit. 
\end{abstract}
\maketitle

\section{Classical enzyme kinetics}

Enzymes are biological catalysts that accelerate chemical reactions
manyfold, without getting consumed in the catalytic process. Several
biological processes involving the conversion of substrates to products,
thus, rely crucially on the catalytic activity of enzymes. Specific
enzymes control and regulate a wide-range of life-sustaining processes
that vary from digestion, metabolism, absorption, blood clotting to
reproduction. While specificity depends on detailed chemical structure
of enzyme proteins, the rate at which enzymes carry out the catalytic
conversion depends less on their chemical structure but more on physical
parameters, including the amounts of enzymes, substrates, temperature,
pH, and so on \cite{key-1}

In 1903, Victor Henri, in his doctoral thesis, studied the rate of
hydrolysis of sucrose into glucose and fructose by the enzyme invertase,
and laid the foundation for the understanding of enzymatic mechanisms
from reaction rates \cite{key-2}. In 1913, Leonor Michaelis and Maud
Menten, building on the work of Victor Henri and many others, introduced
the initial rate method for kinetic analysis. Using data for the initial
rate of hydrolysis of sucrose, a simple reaction mechanism for enzyme
catalyzed reactions, the Michaelis-Menten (MM) mechanism, was proposed
\cite{key-3,key-4,key-5}. 

According to MM mechanism, enzyme E binds with substrate S to form
an enzyme-substrate complex ES, which either dissociates irreversibly
to form product P, regenerating the free enzyme E, or dissociates
reversibly to release the substrate: 

\begin{equation}
E+S\xrightleftharpoons[k_{-1}]{k_{1}}ES\xrightarrow{^{k_{2}}}P+E
\end{equation}
For thermodynamically large numbers of enzymes and substrates, deterministic
mass action kinetics provides the rate equations for the MM mechanism
in terms of the temporal variation of the concentrations of $E$,
$ES$ and $P$. The mean rate of product formation $d[P]/dt$, then,
quantifies the catalytic activity of enzymes through its dependence
on the rate parameters for substrate binding $k_{1}[S]$, substrate
release $k_{-1}$ and product formation $k_{2}$. Assuming a time
scale separation in which the intermediate complex reaches a steady-state
faster than the reactants and products, $d[ES]/dt\approx0$, the mean
rate of product formation yields 

\begin{equation}
V_{ss}=\frac{{V_{max}}[S]}{[S]+K_{M}}
\end{equation}
the celebrated Michaelis-Menten (MM) equation, the steady-state ``enzymatic
velocity'', which has widespread applicability in biochemical catalysis
\cite{key-6}. Here, $K_{M}=\frac{k_{-1}+k_{2}}{k_{1}}$ is the Michaelis
constant and $V_{max}=k_{2}[E_{0}]$ is the maximum velocity at saturating
substrate concentration, with $[E_{0}]$ being the initial enzyme
concentration. The MM equation, in its double reciprocal form, $V_{ss}^{-1}=V_{max}^{-1}+\frac{{K_{M}}}{V_{max}}[S]^{-1}$,
yields a linear variation of $V_{ss}^{-1}$ with $[S]^{-1}$\cite{key-7}.
Kinetic data for the variation of the initial rate of product formation
with $[S]$, when fitted to this linear form, provides a simple way
to estimate the rate parameters of several biochemical reactions.
The hyperbolic dependence of the catalytic rate on substrate concentration,
which the classical MM expression predicts, has had an enormous influence
on the progress of biochemistry. 

\section{Enzyme kinetics at the molecular level}

Enzymatic reactions at the molecular level, however, do not proceed
deterministically. Fluctuations of both quantum mechanical and thermal
origin, termed as molecular noise, are inherent to reactions catalyzed
by individual enzyme molecules. These impart stochasticity to each
step in the MM mechanism, such that neither the lifetime of a given
enzymatic state and nor the state to which it transits can be known
with certainty. The effect of fluctuations, and thus the uncertainty,
diminish progressively with increasing number of enzymes, and vanish
for their macroscopic amounts. It is in the latter limit that the
MM kinetics acquires its deterministic character and the classical
description of enzymatic velocity as the ``mean rate of product formation'',
is sufficient to characterize the catalytic activity of enzymes \cite{key-8}. 

Biochemical catalysis, under physiological conditions, involves enzyme
concentrations that are not thermodynamically large, but vary from
nanomolar to micromolar. The substrates are typically between ten
and ten thousand times more than the number of enzymes. At these low
concentrations, termed as mesocopic, the presence of molecular noise
is inherent to enzymatic reactions \cite{key-9}. Once this fact is
realized, a series of questions arise. Is classical enzymatic velocity,
the MM equation, a valid description of the catalytic activity at
physiologically relevant concentrations of enzymes? How do fluctuations
in the rates of substrate binding, substrate release and product formation
influence the rate of enzyme catalysis? How are uncertainties in the
product formation times measured and characterized? What new information
do these uncertainties carry about the catalytic mechanism that is
lost in going over to the deterministic limit? 

\begin{figure*}
\centering\includegraphics[clip,scale=0.4]{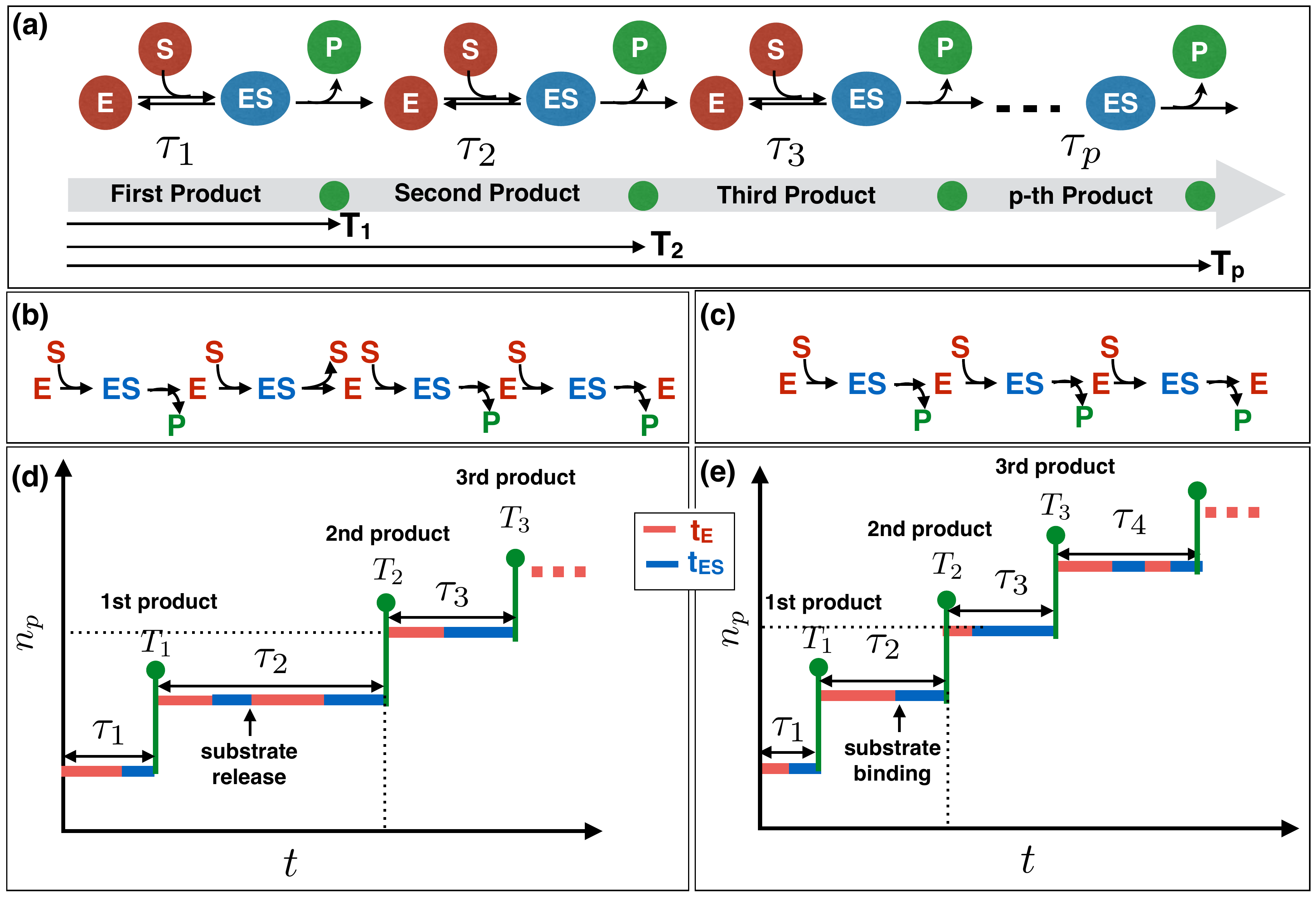}

\caption{Panel (a) shows a schematic of the MM mechanism for a single enzyme
forming products, one by one, in discrete turnover events. Here, $T_{p}$
with $p=1,2,\ldots$ is the $p$-th turnover time and $\tau_{p}=T_{p}-T_{p-1}$
is the waiting time between two consecutive product bursts. Panels
(b) and (c) show two stochastic MM networks, each forming three products
in succession, but following different pathways due to inherent stochasticity
of the MM reaction at the molecular level. Panels (d) and (e) show
two stochastic trajectories, corresponding to (b) and (c), for the
number of products $n_{p}$ formed in time $t$. The lifetimes of
$E$ and $ES$, $t_{E}$ and $t_{ES}$, are indicated by red and blue
colors, respectively. Stochasticity in three consecutive waiting and
turnover times is indicated by solid and dashed lines. \label{fig:mmmechanism}}
\end{figure*}

To address some of these questions, it is important to understand
how single-enzyme kinetic data is collected and analyzed.

\subsection{Single-molecule kinetic data}

In the last two decades, advances in experimental techniques have
finally made it possible to measure, with precision, the effect of
molecular noise in enzyme catalyzed reactions, involving a single
enzyme and numerous substrates \cite{key-10,key-11}. Such techniques
have made it possible to monitor, in real time, the catalytic conversion
of non-fluorescent substrates to fluorescent products, one substrate
at a time, and yield a time series of the turnover times $T_{1},T_{2},\ldots$
for the first, second, $\ldots$ product bursts. The turnover time
series yields an equivalent series of the waiting times $\tau_{1},\tau_{2},\ldots$
between two consecutive product bursts $\tau_{p}=T_{p}-T_{p-1}$,
with turnover number, $p=1,2,\ldots$. 

A schematic of the MM mechanism for a single enzyme forming products
in succession, one product per enzyme turnover, is shown in panel
(a) of Fig. (\ref{fig:mmmechanism}). The waiting and turnover times
for $p=1,2,3,\ldots$ product formation are indicated as $\tau_{p}$
and $T_{p}$, respectively. Each waiting time is the sum of the lifetimes
of $E$ and $ES$ states, $t_{E}$ and $t_{ES}$, $\tau_{p}=t_{E,j}+t_{ES,j}$,
where the subscript index $j=1,2,\ldots$ denotes the number of times
a given state $E$ or$ES$ is visited. 

Due to the presence of molecular noise, the lifetimes of $E$ and
$ES$ states, and hence $\tau_{p}$ and $T_{p}$, are random variables.
Enzyme kinetics at the molecular level, thus, requires a stochastic
description that includes intrinsic fluctuations in the lifetimes
of $E$ and $ES$ states, the waiting times between two consecutive
product turnovers, and the turnover times for $p$-th product formation.
While the duration of time spent in $E$ and $ES$ states remains
unobserved in single-molecule measurements, fluctuations in the lifetimes
of these ``hidden'' states are ``observed'' as fluctuations in the
product formation times $\tau_{p}$ and $T_{p}$. 

To measure fluctuations in $\tau_{p}$, statistical measurements are
carried out, in which waiting times for $p$-th product bursts are
recorded, over and over again, for several stochastic realizations
of the same reaction, under identical experimental conditions. A histogram
of the recorded data is then used to extract probability distributions
of waiting times, $w(\tau_{p})$ and joint distributions of $p$-th
and $q$-th waiting times, $w(\tau_{p},\tau_{q})$. Kinetic data at
the molecular level is, thus, collected as waiting time distributions
$w(\tau_{p})$ and joint distributions $w(\tau_{p},\tau_{q})$, for
given $p,q=1,2,\ldots$ , $N$ and $[S]$. and analyzed in terms of
their moments.

\subsection{Fluctuations in product formation times}

To understand how fluctuations in product formation times can arise,
let us consider two stochastic realizations of the MM mechanism, termed
as stochastic MM networks. Panels $(b)$ and $(c)$ of Fig. (1) illustrate
two stochastic MM networks, each forming three products in succession,
but following different pathways due to the presence of molecular
noise. Stochasticity in these networks arises due to variation in
the lifetimes of $E$ and $ES$ states, and the number of times a
given state is visited. Both these aspects are included in the stochastic
trajectories $(d)$ and (e), corresponding to (b) and (c) respectively,
depicting the number of products $n_{p}$ formed in time $t$. Intrinsic
fluctuations in $t_{E}$ and $t_{ES}$ are indicated as variation
in the length of red and blue lines, and the number of times these
colors switch between each other, before forming a product.

Panels (d) and (e) show that the first waiting time comprises of $\tau_{1}^{(b,c)}=t_{E,1}+t_{ES,1}$
for both (b) and (c). However, given that $t_{E}$ for $(b)$ is longer
than $(c)$, with no variation in $t_{ES}$, implies $\tau_{1}^{(b)}>\tau_{1}^{(c)}$.
This illustrates how stochasticity in the first waiting time can arise
from variation in the lifetime of $E$ state. The waiting times for
the second product burst are given by $\tau_{2}^{(b)}=t_{E,1}+t_{ES,1}+t_{E,2}+t_{ES,2}$
and $\tau_{2}^{(c)}=t_{E,1}+t_{ES,1}$ . These captures stochasticity
in $\tau_{2}$ due to fluctuations in the lifetime of $E$ and $ES$,
and the number of times they switch between each other before forming
a product. Together, they yield fluctuations in the number of products
$n_{p}$ formed in a given time $t$, and the turnover time $T_{p}$
for the $p$-th product formation. These are indicated as dashed lines
in panels (d) and (e). 

\subsection{New statistical measures of fluctuations }

Single enzyme kinetic measurements yield waiting time distributions
for product turnovers. A stochastic reformulation of the MM mechanism,
presented in the next section, allows one to obtain these distributions
theoretically. Enzyme kinetics at the molecular level can, then, be
characterized in terms of the means and variances of these distribution
for given $N$ and $[S]$. The first two moments of $w(\tau)$, for
instance, yields the expectation value $\langle\tau\rangle$, and
variance $\sigma_{\tau}^{2}=\langle\tau^{2}\rangle-\langle\tau\rangle^{2}$
in the waiting time $\tau$. While the former is related to the mean
catalytic rate of a single enzyme, the latter provides a new statistical
measure of intrinsic temporal fluctuations in the catalytic rate \cite{key-12,key-13}.
In particular, the dimensionless variance of the distribution, termed
as the randomness parameter $r=\frac{{\sigma_{\tau}^{2}}}{\langle\tau\rangle^{2}}$,
yields uncertainty in product formation times for a range of substrate
concentrations \cite{key-12,key-13,key-14,key-15,key-16}. 

The correlation, $C_{q}=\langle\delta\tau_{p}\delta\tau_{p+q}\rangle$,
between a waiting time, $\tau_{p}$, and another, $\tau_{p+q}$, $q$
turnovers apart, where $\delta\tau_{p}=\tau_{p}-\langle\tau_{p}\rangle$,
$p=1,2,\cdots$, provides another statistical measure of intrinsic
temporal fluctuations that can be derived from the joint probability
distribution of waiting times \cite{key-9,key-14}. A finite value
of $C_{q}$ indicates that enzymatic turnovers are correlated in time,
and the time duration of first turnover influences the time duration
of subsequent turnovers. The correlations between enzymatic turnovers,
thus, yield a ``molecular memory'' effect in which sequences of waiting
times shorter or longer than the mean are more probable than sequences
uniformly distributed about it \cite{key-9,key-14}. 

The landmark experiments on a single tetrameric enzyme, $\beta$-galactosidase,
which is known to follow the MM equation in bulk amounts, reveal that
the MM equation is violated at the molecular level \cite{key-11}.
This effect due to molecular noise is linked to the simultaneous observation
of $r>1$ and $C_{q}>0$. The magnitude of $r$ and the nature of
decay of $C_{q}$ provide new fundamental constraints on potential
enzymatic mechanisms that can be analyzed theoretically \cite{key-9,key-15,key-16}. 

In the next section, we discuss the relevance of these statistical
measures for the MM mechanism - a network with linear topology comprising
of three reaction steps. We begin by presenting a stochastic reformulation
of the MM kinetics for arbitrary number of enzymes \cite{key-9,key-14}.
An analysis of intrinsic temporal fluctuations, through the probability
distributions of waiting times, is then presented to show how the
statistical measures of intrinsic fluctuations - the randomness parameter
and temporal correlations - carry chemically relevant information
that is lost in passing to the deterministic limit. Specifically,
the turnover kinetics of a single enzyme is compared with mesoscopic
amounts of enzymes to show how deviations from the MM equation are
inherently linked to fluctuations in product formation times.

\begin{figure*}
\centering\includegraphics[clip,scale=0.4]{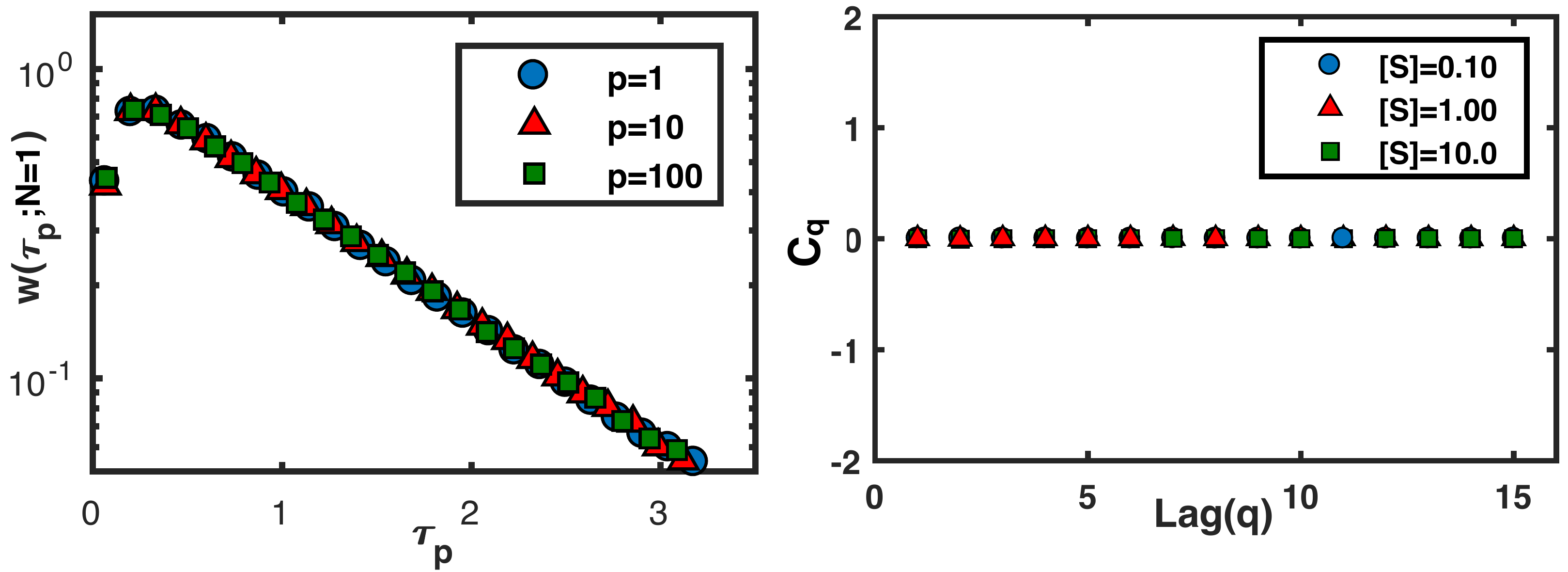}

\caption{Single enzyme kinetics: Left panel shows waiting time distributions
for $N=1$ enzyme for $p=1,10,100$ turnovers. The waiting times,
irrespective of the turnover number, are identically distributed.
Right panel shows that waiting time correlations between, first and
$q=1,2,\ldots$ enzymatic turnovers are identically zero for all $[S]$.
Single enzyme turnover kinetics follows a renewal stochastic process
with independent and identically distributed waiting times. \label{fig:renewal}}
\end{figure*}

\section{Stochastic enzyme kinetics}

Stochastic enzyme kinetics includes inherent fluctuations at the molecule
level by accounting for two important aspects of molecular noise.
These include the discrete integer changes in the number of enzymes,
complexes, products with time, and inherent stochastic character of
each step of the MM mechanism. Both this aspects are included in the
chemical master equation (CME) formalism of stochastic processes,
which provides a general description for the turnover kinetics of
$N$ discrete enzyme molecules \cite{key-17,key-18}. So, while $N=1$
yields ``rate equations'' for a single enzyme \cite{key-12,key-13},
the thermodynamic limit of $N\rightarrow\infty$ reduces the CME to
a set of rate equations, governed by deterministic mass action kinetics
\cite{key-8}. In between the extremes of single enzyme $(N=1)$ and
thermodynamic large $(N\rightarrow\infty)$ limits, the CME describes
the turnover kinetics of mesoscopic enzyme concentrations. 

The CME encodes mechanistic information from which waiting time distributions
can be derived. The latter can be obtained from exact stochastic simulations
of the CME \cite{key-19}. This involves generating a large number
of stochastic trajectories of the MM mechanism, using data for the
number of products versus time to obtain $\tau_{p}$ for $p=1,2,\ldots$,
and histogramming this data to obtain $w(\tau_{p};N)$, for given
$N$ and $[S]$. A similar procedure is followed to obtain the joint
distribution of waiting times. The means and variances of these distributions
provide kinetic measures that yield the enzymatic velocity and fluctuations
in the catalytic rate for $N$ individual enzymes. These, for a single
enzyme, are presented below.

\subsection{Single enzyme kinetics}

To describe the turnover kinetics of a single enzyme, we obtain waiting
time distributions for first, second, third, $\cdots$ product formation,
corresponding to $p=1,2,3,\ldots$ turnovers, using exact stochastic
simulations of the CME \cite{key-9,key-14}. For this we generate
typically $10^{6}$ stochastic trajectories of the MM mechanism for
rate parameters $k_{a}=k_{2}=1$ and $k_{-1}=\frac{{1}}{2}$. The
results are shown in Fig. (\ref{fig:renewal}).

\begin{figure*}
\centering\includegraphics[clip,scale=0.4]{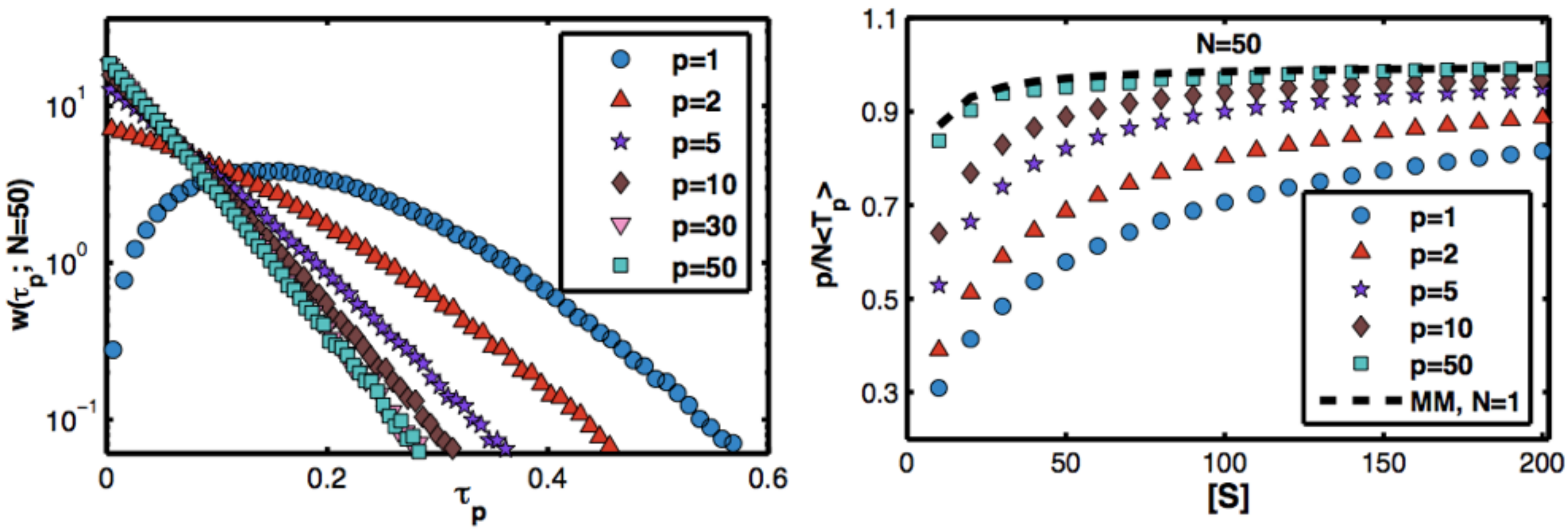}

\caption{Enzyme kinetics at mesoscopic concentrations: Left panel shows waiting
time distributions for $N=50$ enzymes for $p=1,2,5,10,30,50$ turnovers.
The waiting times are non-identically distributed for $p\ll p^{*}$,
the transient regime, but become identically distributed for $p\gg p^{*}$,
the steady-state regime. Here, $p^{*}$ is the critical turnover number
beyond which turnover time correlations decay. Right panel shows that
the scaled enzymatic velocity $p/N\langle T_{p}\rangle$ for $N=50$
enzymes deviates from the MM equation in the transient regime, but
converges to the MM equation in the steady-state regime. \label{fig:non-renewal}}
\end{figure*}

The left panel of Fig. (\ref{fig:renewal}) shows the temporal variation
of waiting time distributions for a single enzyme. The waiting times,
irrespective of the turnover number $p$, are identically distributed.
This implies that the distribution of waiting time between any two
consecutive turnovers, for instance $w(\tau;N=1),$ is sufficient
to describe the turnover kinetics at the single enzyme level, and
specification of the turnover index $p$ is not necessary. The right
panel of Fig. (\ref{fig:renewal}) shows that waiting time correlations
between enzymatic turnovers, $C_{q}$ , are zero for all $q$ and
$[S]$. Together, these results imply that waiting times are independently
and identically distributed. The turnover kinetics of a single enzyme,
thus, follows a renewal stochastic process with $\langle\tau_{p}\rangle=\langle\tau\rangle$
and $\langle T_{p}\rangle=p\langle\tau\rangle$ \cite{key-9}. 

Remarkably, the inverse of the mean waiting time of a single enzyme
exactly recovers the MM equation, $\left<\tau\right>^{-1}=\frac{{k_{2}[S]}}{[S]+K_{M}}$.
Note, that the classical description of the steady-state enzymatic
velocity is in terms of the rate of change of the mean number of products
$V_{ss}=\lim_{t\rightarrow\infty}\frac{{d\langle n_{p}(t)\rangle}}{dt}=\frac{{Nk_{2}[S]}}{[S]+K_{M}}$.
From this, it follows that $V_{ss}=N\langle\tau\rangle^{-1}$. Using
these identities, the classical steady-state enzymatic velocity can
be reinterpreted as

\begin{equation}
\frac{{V_{ss}}}{N}=\langle\tau\rangle^{-1}=\frac{{p}}{\langle T_{p}\rangle}=\frac{{k_{2}[S]}}{[S]+K_{M}}\label{eq:seMM-1}
\end{equation}
This is the single-enzyme analog of the classical MM equation, yielding
a hyperbolic dependence of $\langle\tau\rangle^{-1}$ on substrate
concentration \cite{key-9,key-14}. 

To quantify uncertainty in the waiting time $\tau$, the randomness
parameter, $r=\frac{\left<\tau^{2}\right>-\left<\tau\right>^{2}}{\left<\tau\right>^{2}}$,
is evaluated from the mean and variance of $w(\tau;N=1)$. This yields
$r<1$ for all $[S]$ \cite{key-13}. This result is a specific case
of a formal connection between the randomness parameter and network
topology, which dictates that a reaction mechanism with $n$ sequentially
connected kinetic states, $1\xrightarrow{k}2\xrightarrow{k}3\xrightarrow{k}\cdots\xrightarrow{k}n\xrightarrow{k}n+1$,
with exponentially distributed lifetime of each kinetic state, always
yields $r=\frac{1}{n}$ \cite{key-20}. Here, $n$ represents the
number of rate determining steps in a linear reaction network. Thus,
$r=\frac{1}{n}$ is the minimum amount of uncertainty that can be
captured by the randomness parameter for networks with linear topologies.
A generalization of this to linear networks, in which all (but the
last) nearest neighbor transitions are reversible and occur with arbitrary
rates, $1\rightleftharpoons2\rightleftharpoons3\rightleftharpoons\cdots\rightleftharpoons n\rightarrow1$,
also yields $r<1$. Thus, irrespective of the number of kinetic intermediates,
their connectivity with respect to each other, and rates of transition
between neighboring states, $r\leq1$ for all networks with linear
topologies \cite{key-15,key-16}.

The observation of $r>1$ in recent kinetic measurements on an MM
enzyme, thus, rules out MM networks with linear topologies, and suggest
that MM networks with branches need to be considered \cite{key-13}.

\subsection{Enzyme kinetics at mesoscopic concentrations}

\begin{figure*}
\centering\includegraphics[clip,scale=0.4]{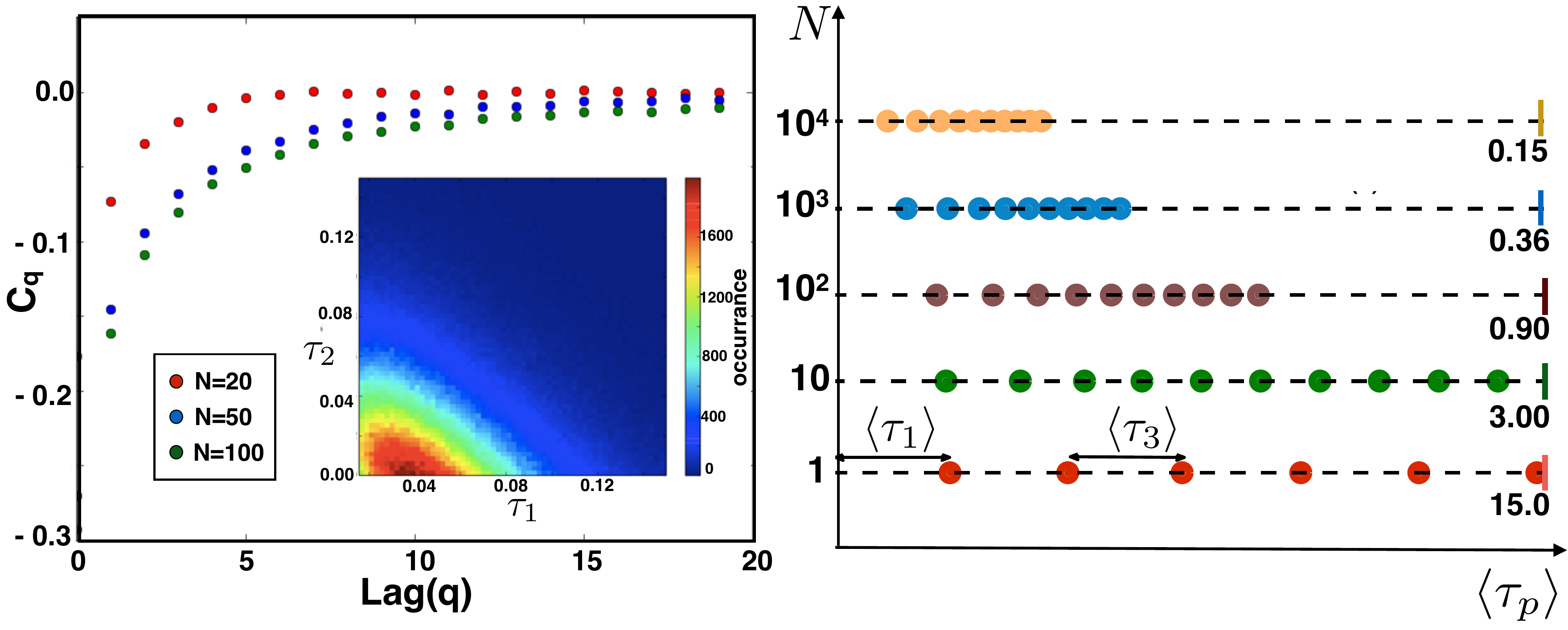}

\caption{Left panel shows temporal correlations between $\tau_{1}$ and $\tau_{q}$
with $q=1,2,\ldots$ waiting times. The waiting times are anticorrelated,
where a short (or long) first waiting time is more likely to be followed
by a long (or short) second waiting time (inset). This memory effect,
though stronger and short lived for smaller $N$, becomes weaker and
long lived for larger $N$. Right panel shows the variation of $\langle\tau_{p}\rangle$,
for first ten turnovers $p=1,2,..10$ with increasing N. The discrete
turnover events acquire a deterministic character, with the increase
in $N$, as waiting times become infinitesimally short and products
appear to form continuously in time. \label{fig:corr}}
\end{figure*}

In between the classical limit of thermodynamically large number of
enzymes and a single enzyme, lies mesoscopic number of enzymes. To
obtain waiting time distributions for mesoscopic concentrations, we
carry out exact stochastic simulations of the CME \cite{key-9,key-14}.
The results are presented in Figs. (\ref{fig:non-renewal}) and (\ref{fig:corr}). 

The left panel of Fig. (\ref{fig:non-renewal}) shows the temporal
variation of waiting time distributions for increasing turnover numbers.
The right panel shows, for the same turnover numbers, the variation
of the enzymatic velocity of $N$ discrete enzymes, $\frac{{p}}{N\langle T_{p}\rangle}$,
with substrate concentration. Interestingly, for turnover numbers
much less than the critical turnover $p^{*}$ , $p\ll p^{*}$, the
waiting times are non-identically distributed and the MM equation
is not obeyed. For $p\gg p^{*}$, the renewal turnover statistics
and the MM equation are asymptotically recovered. 

To understand this, the left panel of Fig. (\ref{fig:corr}) shows
the variation of waiting time correlations between enzymatic turnovers
$C_{q}$ as a function of $q=1,2,\ldots$. For mesoscopic amounts
of enzymes, temporal correlations between enzymatic turnovers, though
appreciable for $p\ll p^{*}$, become negligible for $p\gg p^{*}$.
Here, $p^{*}$ is the critical turnover number beyond with turnover
time correlations decay. 

Together, these results imply that the critical turnover number demarcates
a transient regime $p\ll p^{*}$ from a steady state regime $p\gg p^{*}$
\cite{key-14}. In the transient regime, turnover kinetics is non-renewal,
where waiting times are non-independent and non-identically distributed,
and the MM equation is violated. In the steady-state regime, turnover
kinetic acquire a renewal character and the MM equation is exactly
recovered. The non-renewal nature of the turnover kinetics implies
that $\langle T_{p}\rangle\neq p\langle\tau_{p}\rangle$. For mesoscopic
amounts of enzymes, then, the steady-state enzymatic velocity can
be reinterpreted as

\begin{equation}
V_{ss}=\lim_{p\rightarrow\infty}\frac{{p}}{\langle T_{p}\rangle}=\frac{{Nk_{2}[S]}}{[S]+K_{M}}.\label{eq:mes-MM}
\end{equation}
Note that the above expression is only valid for $p\gg p^{*}$, where
$C_{q}=0$. 

For the MM mechanism, the waiting times between enzymatic turnovers
are anti-correlated, $C_{q}<0$. This implies that a long (or short)
first waiting time (compared to its mean value) is more likely to
be followed by a short (or long) second waiting time. This memory
effect, is shown as the heat map of joint probability distribution
$w(\tau_{1},\tau_{2})$ in Fig. (\ref{fig:corr}) inset. 

\subsection{The classical limit from molecular perspective}

To understand how the classical limit of enzyme kinetics can emerge
from molecular kinetics, we show in the right panel of Fig. (\ref{fig:corr}),
the variation of $\langle\tau_{p}\rangle$ for first ten product turnovers,
as a function of increasing $N$. For $N=1$, since $C_{q}=0$, the
mean waiting times are identical $\langle\tau_{p}\rangle=\langle\tau\rangle$,
and the MM equation, $\left<\tau\right>^{-1}=\frac{{k_{2}[S]}}{[S]+K_{M}}$,
is always obeyed. 

For $N>1$, there emerge two kinetic regimes - a transient regime
for $p\ll p^{*}$ , and a steady-state regime for $p\gg p^{*}$. In
the transient regime, non-identical mean waiting times $\langle\tau_{p}\rangle$
become progressively shorter with increasing $N$ and $p$. In the
steady-state regime, similarly, there is an $N$-fold decrease in
the mean waiting time $\langle\tau\rangle$. The discrete turnover
events acquire a deterministic character, with the increase in $N$,
as waiting times between consecutive products become shorter, and
the number of products in a small time interval become progressively
larger. 

In the limit of macroscopic amounts of enzymes $N\gg1$, thus, waiting
times become infinitesimally small, products appear to form continuously
in time, and the transient regime becomes too short to be experimentally
accessible. In this classical limit, kinetic data for the ``initial''
mean rate of product formation yields the steady-state enzymatic velocity.
From the molecular perspective, the absence of temporal correlations
between enzymatic turnovers in the steady-state regime, implies that
the MM equation is obeyed. 

\section{Conclusions}

In this mini-review, a stochastic reformulation of the MM kinetics
for a single and mesoscopic amounts of enzyme(s) has been presented.
An analysis of intrinsic temporal fluctuations, through the probability
distributions of waiting times, shows that the MM equation is always
violated in the transient regime. This regime, characterized by temporal
correlations between enzymatic turnovers, while absent for a single
enzyme, is unattainable in the classical limit. In both these limits,
thus, the MM equation, is exactly obeyed. 

At mesoscopic concentrations, the transient regime is observably large,
and temporal correlations between enzymatic turnovers yields a molecular
memory effect in which the time duration of the first turnover influences
the time duration of subsequent turnovers. In the presence of molecular
memory, the MM equation is not obeyed at the molecular level. This
effect, due to molecular cooperativity, leads to a slowing down of
the MM kinetics in the transient regime. The critical turnover time
beyond which the correlation decays and the molecular memory fades
marks the time scale for the crossover from a transient regime where
the MM equation is violated, to a a steady-state regime where the
MM equation is recovered exactly at the molecular level. 

From these general results, we conclude that for mesoscopic amounts
of enzymes, there emerges a transient regime in which enzymatic turnovers
are correlated in time, and the the MM equation, derived using time
scale separation, is always violated.

\end{document}